\documentclass[prb,twocolumn,showpacs]{revtex4}
\usepackage{graphicx}% Include figure files
\usepackage{graphics}
\usepackage{dcolumn}% Align table columns on decimal point
\usepackage{bm}% bold math
\newcommand{\beq}{\begin{equation}}
\newcommand{\eeq}{\end{equation}}
\newcommand{\beqnar}{\begin{eqnarray}}
\newcommand{\eeqnar}{\end{eqnarray}}
\newcommand{\bfig}{\begin{figure}}
\newcommand{\efig}{\end{figure}}

\begin{document}
\title{Negative differential resistance in molecular junctions: application to graphene ribbon junctions}

\author{Hosein Cheraghchi$^{1}$, Keivan Esfarjani$^{2}$}

\affiliation{$^{1}$School of Physics, Damghan University of basic
sciences, Damghan, IRAN \\ $^{2}$Department of Physics,
University of California, Santa Cruz, CA 95064}
\email{cheraghchi@dubs.ac.ir}
\date{\today}
\begin{abstract}
Using self-consistent calculations based on Non-Equilibrium
Green's Function (NEGF) formalism, the origin of negative
differential resistance (NDR) in molecular junctions and quantum
wires is investigated. Coupling of the molecule to electrodes
becomes asymmetric at high bias due to asymmetry between its
highest occupied molecular orbital (HOMO) and lowest unoccupied
molecular orbital (LUMO) levels. This causes appearance of an
asymmetric potential profile due to a depletion of charge and
reduction of screening near the source electrode. With increasing
bias, this sharp potential drop leads to an enhanced localization
of the HOMO and LUMO states in different parts of the system. The
reduction in overlap, caused by localization, results in a
significant reduction in the transmission coefficient and current
with increasing bias. An atomic chain connected to two Graphene
ribbons was investigated to illustrate these effects. For a chain
substituting a molecule, an even-odd effect is also observed in
the NDR characteristics.
\end{abstract}
\pacs{73.23.-b,73.63.-b} \keywords{} \maketitle
\section{Introduction} Negative differential resistance (NDR)
was first observed by Esaki in diodes\cite{esaki}, where occupied
states on one side become aligned with the gap of other side as
the voltage is increased. Current reduction also occurs when the
position of the resonant states of the molecule move within the
gap of one of the contacts \cite{esaki-rtd,nanolett1} as in
resonant tunneling diodes. In metallic carbon nanotube junctions
\cite{farajian}, it was found that the reduction of the current
is due to a mismatch in the symmetry of the incoming and outgoing
wavefunctions of the same energy. Another work\cite{orb-match} on
the I-V characteristic of CoPc on gold has also associated the
NDR effect with lack of orbital matching between Ni tip and Co
atom. Another origin was explained in STM
measurements\cite{Avouris,STM}. In this case, narrow peaks in the
local density of states (LDOS) of an atomic scale tip sweep past
the LDOS of an adsorbed molecule as the bias voltage is increased.

More recently, more instances of NDR were
observed\cite{nanolet,Louie_NT,orb-match} or predicted
\cite{NDRgraphene,vanhove,kirczenow,korean,datta-spin,Lang} in
molecular devices. In the case of potential barriers in 2D
Graphene sheets\cite{NDRgraphene}, the effect was due to the
linear dispersion of (massless Dirac) electrons which show a gap
in their transmission across the barrier. In Ref.
\onlinecite{vanhove} it was due to the presence of Van Hove
singularities in the DOS of the 1D electrodes regardless of the
type of the contact. This latter explanation is related and
similar to that of Refs.[{\onlinecite{Avouris,Lang,STM}}] which
involves sharp features in the LDOS. In these cases, however, the
general conditions necessary for the observation of the effect
were not clearly elucidated. Sharp features in the LDOS can lead
to NDR \cite{Lang,Guo}, but it is not a sufficient condition for
the observation of NDR, as a reduction in {\it spatial overlap} of
those states is also needed.

The current in nanoscale devices is given by the Landauer formula
(see eq. \ref{current}) which involves the transmission
coefficient given by the product of the local density of states
(LDOS) of the left and right electrodes by the off-diagonal
matrix elements of the Green's function (GF) connecting the left
electrode to the right one (see eq.\ref{transmission}). A
reduction in the current is caused by a lowering of either term
in the transmission coefficient. While NDR in some devices is
caused by a lowering of the matrix element of the
GF\cite{ndr-gf}, in some other cases it is caused by a reduction
in the product LDOS within the energy integration
window\cite{Lang,Avouris,STM,orb-match,korean,Guo}.

In this paper we explain the reason for occurrence of sharp
features in LDOS, and also emphasize that charging effects play
an enhancing role in producing NDR in the I-V characteristics of
nano-junctions. A large bias causes charge depletion, an
asymmetric potential profile, and asymmetric coupling even in a
symmetric structure, resulting in a stronger localization of
states on different parts of the system, thereby reducing
transmission and current.

 We consider an atomic carbon chain between two
graphene tips as a nano-junction (Fig. \ref{graphene-chain}),
albeit all results are generalizable to other types of
nano-junctions. Weak contacts between tips and the chain/molecule
which usually occur in experiments involving break or molecular
junctions, are necessary for causing localized states within the
molecular region and observation of NDR. So, we adopt a model in
which hoppings to leads are smaller than intramolecular or
intralead hoppings. We claim that in molecular junctions where
NDR is observed, localization of electronic states within the bias
energy window is the dominant cause of reduction in current. The
weak bond can play the role of a barrier to localize states
within or near the molecule. The purpose of our model is not to
make quantitative predictions, but just to illustrate the NDR
mechanism using a simple enough model. Given the small size of
contact we assume that transport at high bias is mostly coherent
and dissipation due to electron-phonon interactions occurs mainly
in the drain.

After presenting Hamiltonian of the system in section II, we will
introduce the formalism and method used to handle the
electrostatics of the problem in the section III. In the
Appendix, electrostatic potential calculated by this method is
compared with two other methods. We are going over the general
formalism used for the calculation of non-linear transport
characteristics in section IV. The responsible for current
reduction in an atomic chain between two graphene tips which is
known to be localization of states induced by charging effects
will be presented in the section V.

\section{Model}

%---------------- Fig.1 ------------------------
\bfig
\includegraphics [width=7 cm] {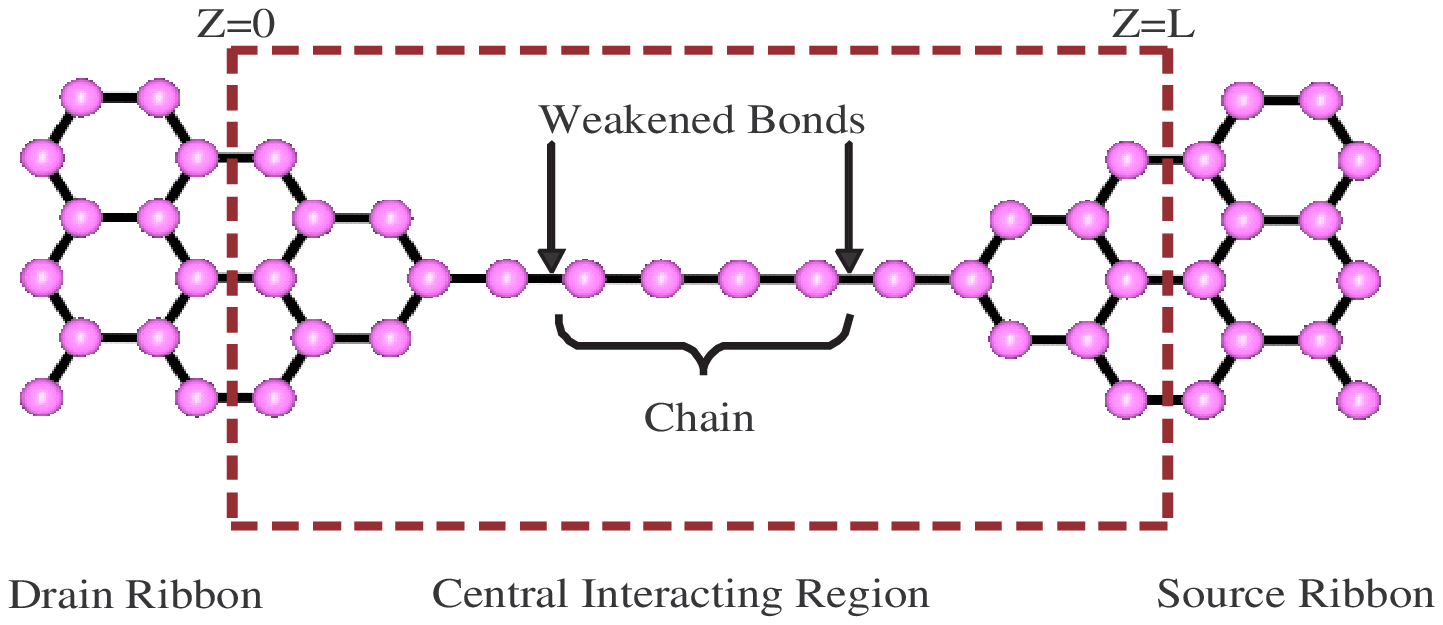}
\caption{(Color online) Two Graphene tips connected with a weakened bond to a
carbon chain. Tips have sharp structures. The weakened bond is
considered to be $0.3$ of the normal hopping of the C-C bond.
The central interacting region is shown with the dashed rectangle.}
\label{graphene-chain} \efig
%------------------------------------------------

The single electron Hamiltonian of the central system (C) including the molecule is

\beq%\begin{array}{r}
H_C=\sum_{i \in C} %=1}^{N}
[\varepsilon_i+u^{ext}_{i}
+%\sum_{j} =1}^{N} V_{ij}\delta n_j
W_i] c^{\dagger}_{i}c_{i}
 +\sum_{<ij>}t (c^{\dagger}_{i}c_{j}+c_{i}c^{\dagger}_{j})
% \end{array}
\label{hamiltonian}\eeq

where  $c^{\dagger}_{i}$ and $c_{i}$ are respectively the electron creation and
annihilation operators on site $i$ of $C$, and  $t$ is
the hopping energy between nearest neighbor atoms.
One $\pi$ orbital per site is considered for this
system. Under an applied bias, the solution to Poisson's equation
is the sum of the solution to {\it Laplace} with symmetric
boundary conditions on the electrodes $V(z=0)=-V/2$ and
$V(z=L)=V/2$ (this is denoted by $u^{ext}_{i}$), and the {\it
solution to Poisson} with boundary condition
$V(z=0)=V(z=L)=0$ at both ends (this is denoted by $W_i=\sum_j V_{ij}\delta
n_j$). The sum $u^{ext} + W $ clearly satisfies Poisson
equation and the proper boundary conditions. Here $V_{ij}$ is the
electrostatic Green's function calculated by the method of
images, and $\delta n_j=n_j-n_j^0$ is the change in the
self-consistent charge $n_j$ from its initial equilibrium
zero-bias value.

It should be noted that parts of electrodes (here also called as
"tips") have been incorporated inside the interacting central
region as there is always some potential drop beyond the contact
of the electrodes with the central ``molecule".

\section{Electrostatic Green's Function}

The electrostatic potential is determined by both the direct
interaction of electrons with each other and the indirect one via
image charges. The image charges induced by electrons within the
electrodes, strongly depend on the spatial configuration of the
electrodes and the contact atoms. For the simplicity of
calculations, it is usual to consider the electrodes as two
infinite planes perpendicular to the molecule\cite{Sablikov}.
These planes are located on the contacts.

It is supposed that electrodes are a perfect metal with good
screening properties, and that at their boundary the potential
can be considered as a constant so that Dirichlet boundary
conditions can be applied there. In this case, the potential drop
occurs within the central part of the sample, which we call
``molecule", although, strictly speaking this central region is
taken to be larger than the molecule itself as there is always
some potential drop at the contact of the electrodes with the
central ``molecule".

It should be mentioned that the 3-dimensional Poisson equation
needs to be solved in order to find the correct potential profile
along the molecule. Indeed the electric field lines are not
necessarily straight lines, and a 1D solution would be incorrect.
So the Coulomb Kernel needs to be more like the 3-dimensional
$1/|r-r'|$ rather than the 1-dimensional $|r-r'|$.

As there is a finite charging energy when the two electrons are
on the same site, there should be no divergence in the kernel,
and the onsite Coulomb repulsion has been modeled by the
so-called ``Hubbard" parameter $U_H$, which could also contain
exchange and correlation effects if appropriately chosen.
However, image charges potential lowers the potential on one site
from its initial value $U_H$.

In this article, the Ohno-Klopmann (OK) model \cite{ok} has been
adopted for the Coulombic function $U$:

\beq
U(\vec{r}_i,\vec{r}_j)=\frac{1}{\sqrt{\mid\vec{r}_i-\vec{r}_j\mid^2+U_{H}^{-2}}}
\label{OK}
 \eeq
It has the correct limits for both large and small inter-particle
distance $\vec{r}_i-\vec{r}_j$. It has the advantage of including
onsite correlations through the Hubbard-like parameter $U_H$.

In the literature \cite{Jackson}, there exists an exact Dirichlet
Green's function for a point charge or a distribution of charges
between parallel conducting planes held at zero potential. The
planes are located at z=0 and z=L.  Using this Green function, we
present the following exact form which is appropriate for the
kernel of Ohno-Klopmann model (Eq.(\ref{OK})).

\beq V(x,y,z;x^{'},y^{'},z^{'})=2\int_0^{\infty} dk J_0(\alpha
k)f(k,z_{<},z_{>})
 \label{exact}\eeq
 where
 \beq
f(k,z_<,z_>)=\frac{\sinh(k z_<)\sinh(k (L-z_>))}{\sinh(k L)}
\label{fofk}
 \eeq
\beq
 \alpha=\sqrt{(x-x^{'})^2+(y-y^{'})^2 + U_H^{-2}}
\eeq
The asymptotic behavior of the function $f(k)$ in Eq.(\ref{fofk}) is
as follows:

\beq \lim_{k\rightarrow\infty} f(k)\rightarrow 0.5 e^{-k
(z_>-z_<)} \rightarrow\left\{
\begin{array}{c c}
0.5&\,\,\ z_{<}=z_{>} \\
 0 & \,\,\ z_< \neq z_>
\end{array} \right.{ \label{func}}
 \eeq
Moreover, $f(k)$ goes to zero when $k\rightarrow0$. Since at
$z_{<}=z_{>}$, the function of $f(k)$ will be a constant for
$k\gg 1/z_<$, the integration with infinite range can be
converted to a limited range integration .

 \beq
V(z_<=z_>)=\frac{1}{\alpha}-\int_0^{k_0}(1-2f(k))J_0(\alpha k) dk
 \label{zeqzp}\eeq
where $f(k_0)=0.5$. The value of $k_0$ in nanotubes and graphenes
used here is about 100. This value depends on the distances
between atoms of a molecule and also on the distances between two
boundary planes ($L$). In case of on site electrostatic potential
($x=x^{'};y=y^{'}$), the first term of Eq.(\ref{zeqzp}) is the
Hubbard energy. However, a subtraction term which depends on the
distances between atoms and $L$, lowers the Hubbard energy from
$U_H$. This term is the image charges potential which was
considered in the variational method, too (Appendix.B). The value
of the semi-empirical Hubbard term for carbon\cite{sctb}, is
about 10 eV=0.37 a.u. So $U_H^{-1}\cong 2.72$ whereas the typical
bond length is of the order of 1.4 \AA=2.6 a.u.

In the Appendix (A,B), we compare this method (namely the exact
method) with two other methods so-called the variational and
image charges method.

\section{Calculation of Charge and Current} \label{negf}

The charge is obtained using the NEGF formalism
\cite{Liang,Datta}. The electrodes electrochemical potentials and
the fermi functions are shown by $\mu_{L,R}$ and $f_{L,R}$,
respectively. The retarded Green's function matrix is: \beq
G(Z)=[ZI-H-\Sigma_L^r-\Sigma_R^r]^{-1} \label{Green}\eeq

where $Z=E+i\eta$ is a complex variable whose real part is energy
and $\eta\rightarrow0^+$. "$I$" is the unit matrix. $H$ is the
molecule Hamiltonian defined by Eq.(\ref{hamiltonian}) in the
tight-binding approach. $\Sigma_{L/R}^r$ are the retarded
self-energies arising from scattering by the left/right
semi-infinite electrodes. These self-energies depend on space
configuration of the electrodes and the quality of the
electrode-molecule couplings. We have to obtain the surface
Green's function of semi-infinite electrodes $g_{p}(E)$ in order
to determine the self-energy. The Lopez-Sancho's method
\cite{Munoz} has been used to calculate the surface Green's
function. The retarded self-energies are given by:

 \beq
\Sigma^r_{p}=\tau_p^T g_p^r(E) \tau_p \,\,\,\;\,\,\ p\equiv L/R
\label{self-energies} \eeq

 where $\tau_p$ is the coupling matrix between the electrodes and the
 molecule \cite{Datta}. Since the hopping terms are short-ranged, most elements
 of the coupling matrix are zero. Broadening of the molecule energy levels due to
 attachment to the electrodes is related to the self-energies as:

 \beq
\Gamma_p=i[\Sigma_p^r-\Sigma_p^a] = 2 \pi \tau_p^T LDOS(p,E)
\tau_p \eeq
 Note that the broadenings are proportional to the
local density of states at the connecting sites to the
electrodes. It should be noted that in this paper transport is
assumed to be coherent. The charge density is the sum of two
separate parts coming from equilibrium and non-equilibrium
charges. Since the voltage division is symmetric on the
electrodes, the equilibrium charge $n_{eq}$ is calculated from
the retarded Green's function as:

 \beq
n^{eq}_i=\frac{-1}{\pi}\int_{-\infty}^{\mu_0-V/2}Im[G^r_{ii}(E)]dE
\label{eqcharge} \eeq

where $\mu_0=\mu_R=\mu_L$. The initial charge $n_i^0$ is
calculated by the above integration in zero bias. In the
non-equilibrium situation, the lesser Green's function $-iG^<(E)$
represents the occupation number in the presence of the two
electrodes subject to a bias. The non-equilibrium charge
$n_{non-eq}$ is determined in the presence of an external bias
$V$.

 \beq
n^{non-eq}_i=\frac{1}{2\pi}\int_{\mu_0-V/2}^{\mu_0+V/2}[-iG_{ii}^<(E)]dE
\eeq

It can be simply shown that in the coherent regime the lesser
Green's function is determined by the retarded Green's function
(Eq.(\ref{Green})).

 \beq
n^{non-eq}_i=\frac{1}{2\pi}\int_{\mu_0-V/2}^{\mu_0+V/2}[G^r(\Gamma_L
f_L+\Gamma_R f_R)G^a]_{ii}dE \label{noneqcharge} \eeq

where $f_p=1/[1+\exp(\frac{E-\mu_p}{k_B T})]$ shows the fermi
function of the electrodes. Finally, both parts of the charge are
summed to give the total charge:

 \beq n=n^{eq}+n^{non-eq}\label{charge} \eeq

Since the molecular Hamiltonian itself depends on the electron
density, one needs to do a self-consistent process. The
self-consistent algorithm follows these steps. At the first step,
the left and right self-energies in Eq.(\ref{self-energies}) are
calculated once before the self-consistent loop. In the second
step, the Hamiltonian is set using a guess input charge.

The calculation of charges in
Eqs.(\ref{eqcharge},\ref{noneqcharge}) is a hard step as it needs
to be well converged. The new and old charges can be mixed with
each other by using linear mixing or Broyden's method
\cite{broyden-ohno}. Using the mixed charge, this process will
start from the first step and continue till convergence is
achieved. Finally, having the self-consistent charge and
potential profiles, the current passing through the molecule is
calculated by the Landauer formula \cite{Datta}.

\beqnar I(V)&=&\frac{2e}{h}\int_{-\infty}^{\infty}\,dE \,T(E,V)\,
[f_R(E)-f_L(E)] \nonumber \\
&=&\frac{2e}{h}\int_{\mu_0-V/2}^{\mu_0+V/2}\,dE \,T(E,V)\
\label{current}\eeqnar where the second expression is written for
zero temperature. The transmission coefficient $T(E,V)$ is
defined as:
 \beq T={\rm Tr}[G^r \Gamma_R G^a \Gamma_L] \propto
{\rm LDOS}(L) {\rm LDOS}(R) |G_{LR}|^2 \label{transmission}\eeq

The integral evaluation for charge density in
Eqs.(\ref{eqcharge},\ref{noneqcharge}) has to reach a reasonable
accuracy. The speed of the convergence process depends strongly
on the accuracy of the integration process. For weak couplings,
the van Hove singularities in the density of states (DOS) will
make it tremendously difficult to integrate the DOS along the
real axis with desired accuracy. Indeed, the singularities arising
from the poles of the Green's function are close to the real
axis. However, in the complex energy plane, the DOS along the
complex contour away from the real axis is very smooth
\cite{Taylor}. The resultant formula for a contour integration of
the equilibrium charge is:

\begin{eqnarray}
n^{eq}_i&=&\frac{\rho}{\pi}\int_0^{\pi}Re[G_{ii}(z_0+\rho e^{i
\theta})e^{i \theta}]d\theta \nonumber \\
\\
\rho&=&\frac{\mu_0-V/2-E_{min}}{2};z_0=\frac{\mu_0-V/2+E_{min}}{2}
\end{eqnarray}

where $E_{min}$ is chosen to be lower than the lowest eigenvalue
of $H$.

%-------Fig. 2 -------------------------
 \bfig
\includegraphics [width=7 cm] {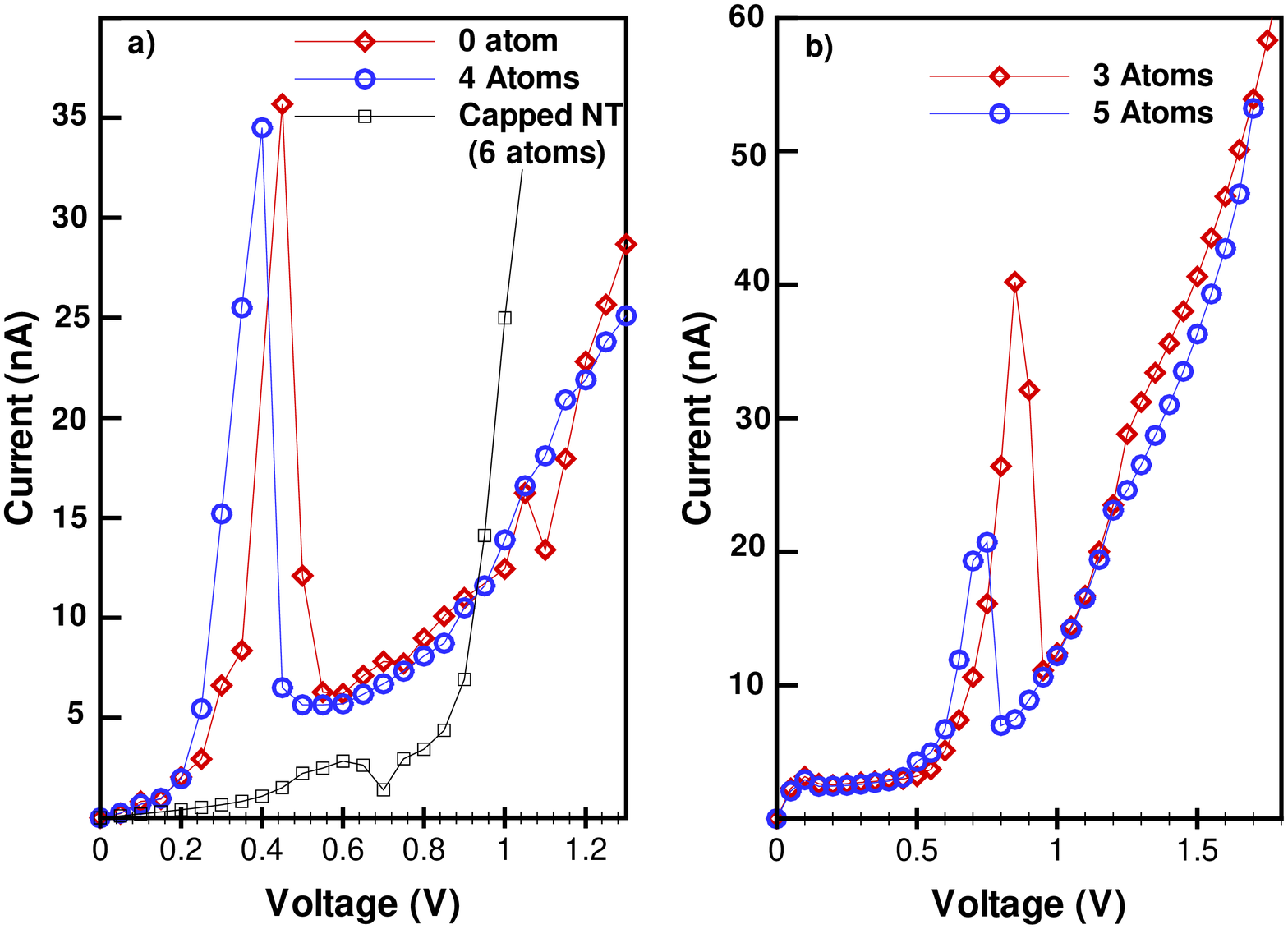}
\caption{(Color online) I-V curves for two graphene tips connected to the chain
with a) even and b) odd chains. The hopping of the weakened bond
is 0.3 times that of the intra-chain hopping
($t$). In case of "0 atom" where two tips are facing each other
with no chain in between, the hopping integral is equal to 0.1
$t$. NDR is also observed in two (5,5) capped nanotubes (NTs) with
a 6 atoms chain in between. Current through NT system is 50 times
larger than shown.} \label{IV-GR} \efig
%----------------------------------------

\section{Results}

Fig.(\ref{IV-GR})  shows the NDR phenomenon in the I-V curves of
odd and even length chains located between two graphene tips.
%There is an odd-even parity in the I-V curve.
To show that NDR is also obtained with gapless leads, we have also made
calculations for a (5,5) carbon nanotube
and still observed a reduction in the
current due to localization of states at the caps of the tubes at
high bias. Details will be reported elsewhere.
The NDR threshold voltage for odd length chains is higher than
that for even chains. The origin of this difference can be traced
back to the distance of those levels which play a role in the
observation of NDR from the Fermi level. For odd chains, the
state at the Fermi level is an extended state over the length of
the chain \cite{cheraghchi}, whereas even chains have a gap at
the Fermi level. Therefore typically a twice larger bias is
needed to observe NDR in odd chains compared to even chains of
similar length.

%------- Fig. 3 -----------
\bfig \includegraphics [width=7 cm]{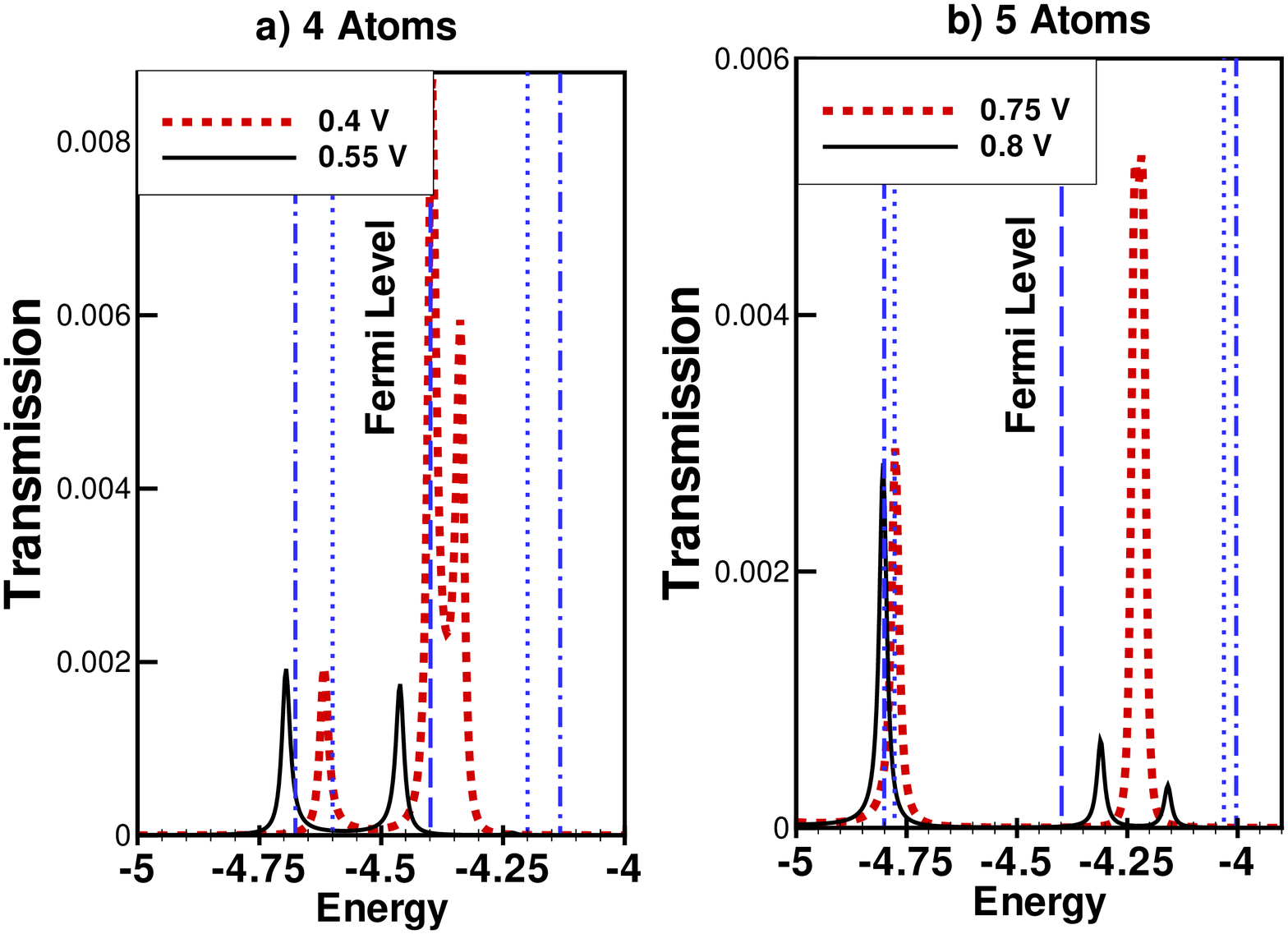}
\caption{(Color online) Transmission coefficient through a chain connected to
two graphene tips. Transmission is plotted for a chain with a) 4
atoms (even chain) b) 5 atoms (odd chain) at peak and valley
voltages. Vertical dashed, dotted and dash-dotted lines identify
the Fermi level, integration windows at current-peak voltage and
current-valley voltage, respectively. A large reduction in the
transmission can be noticed at higher voltage. Transmission
through other odd/even chains have similar features.}
\label{Transmission_GR}\efig
%------------------------------------------

To understand the origin of NDR in this system, in
Fig. \ref{Transmission_GR}  we compare the transmission
coefficients at the current peak and valley voltages.
As one can see from the figure,
there is a large reduction in the transmission of the resonant states
when the bias is increased.
We will show that the reason for this
can be traced back to a loss of LDOS overlap of the left contact with
the right one.

%------ Fig. 4 -----------
\bfig
\includegraphics [width=7. cm] {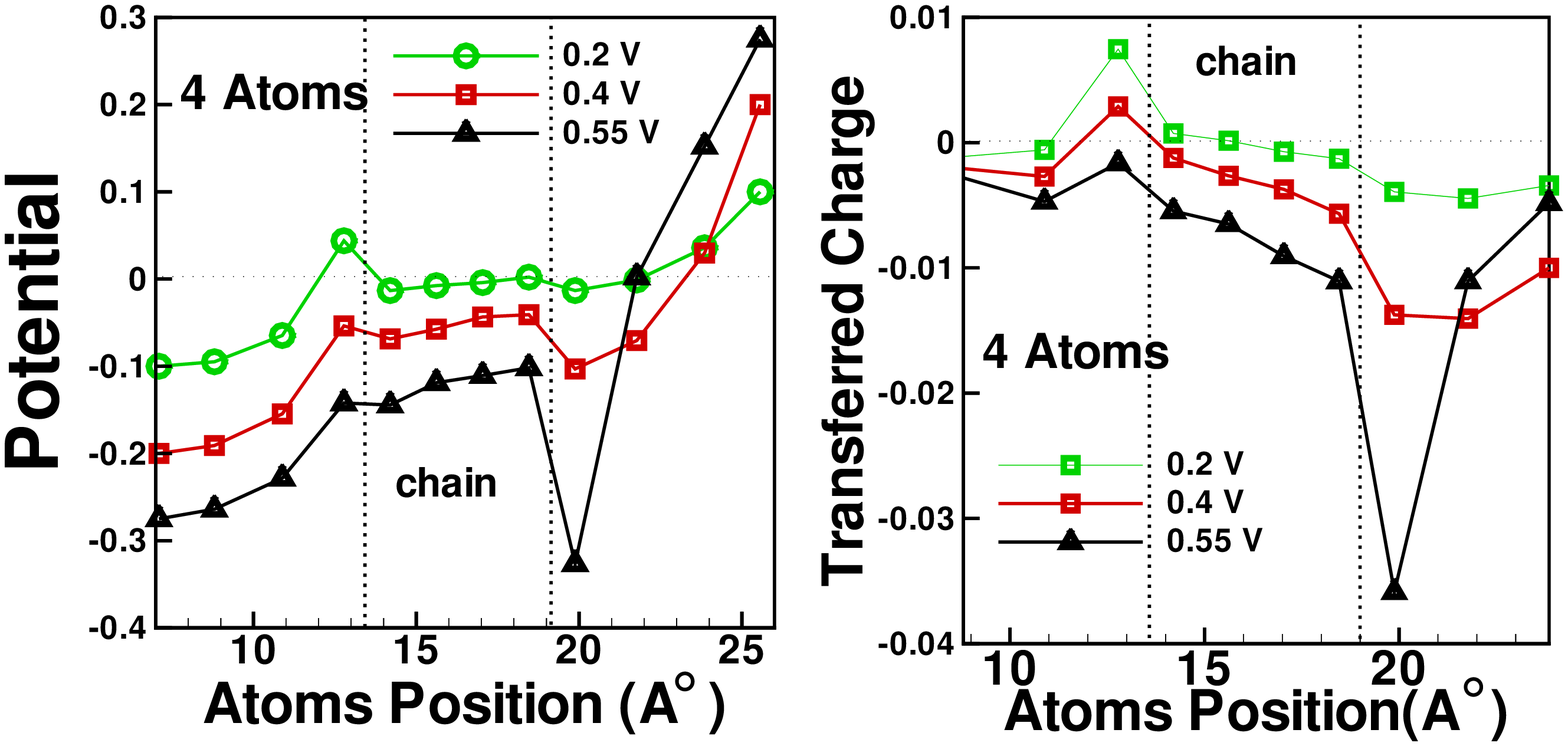}
\caption{(Color online) Potential and transferred charge $(n_i-n_i^0)$ for a
chain with 4 atoms between two Graphene tips. Profiles for three
voltages are plotted; a small voltage, and voltages which
correspond to the current peak and valley. Potential and charge
has been averaged on each Graphene layer. Source is on the right and drain on the left.}
\label{Charge_potential-GRPH} \efig
%-------------------------

In Fig. \ref{Charge_potential-GRPH}, the electrostatic potential
energy and transferred charge ($\delta n=n-n_0$) profiles are
plotted for different biases. These distributions are obtained
for a small voltage (0.2 V) and voltages of the peak and valley
of the current. In the linear regime, potential is nearly
symmetric. However, by increasing the bias, some charge is
depleted from the source, thereby weakening the effect of
screening and enhancing the potential drop further at the source.
The asymmetry in the voltage drop can be understood in the
following way. The transferred charge between electrodes and
molecule depends on the quantum capacitance of the molecule.
Quantum capacitance increases with the surface density of states
at the source or drain electrochemical potentials. Fig.
\ref{SDOS} (a,b) shows that $LDOS(E_F+V/2)$ on the surface layer
of the source side is much smaller than $LDOS(E_F-V/2)$ on the
drain side. Due to its capacitive coupling with the drain, one
state (see Fig. \ref{SDOS}(a)) which is localized on the drain
side of the molecule is pinned at $E_F-V/2$. So $LDOS(E_F-V/2)$
remains large as the bias is increased, while $LDOS(E_F+V/2)$
gradually decreases when the resonant states in Fig.
\ref{SDOS}(b) move away from $E_F+V/2$. This asymmetry in LDOS
translates into an asymmetry in the couplings of the central
region to leads, even though there geometric symmetry is enforced.
On the side with weaker coupling (source side in our case)
screening would be less effective and potential drop more
pronounced. Therefore essentially the asymmetry at large biases
develops due to the asymmetry in the distribution of molecular
states around the Fermi level. This phenomenon is expected to be
universal in molecular double junctions with weak couplings.
Another consequence of the effective weakening of the couplings to
the leads is the sharpening of the molecular states. States near
the weak coupling will have narrower peaks at high bias. This is
a signature of their enhanced localization.
%-------------- Fig. 5 ---------------
\bfig
\includegraphics [width=7 cm] {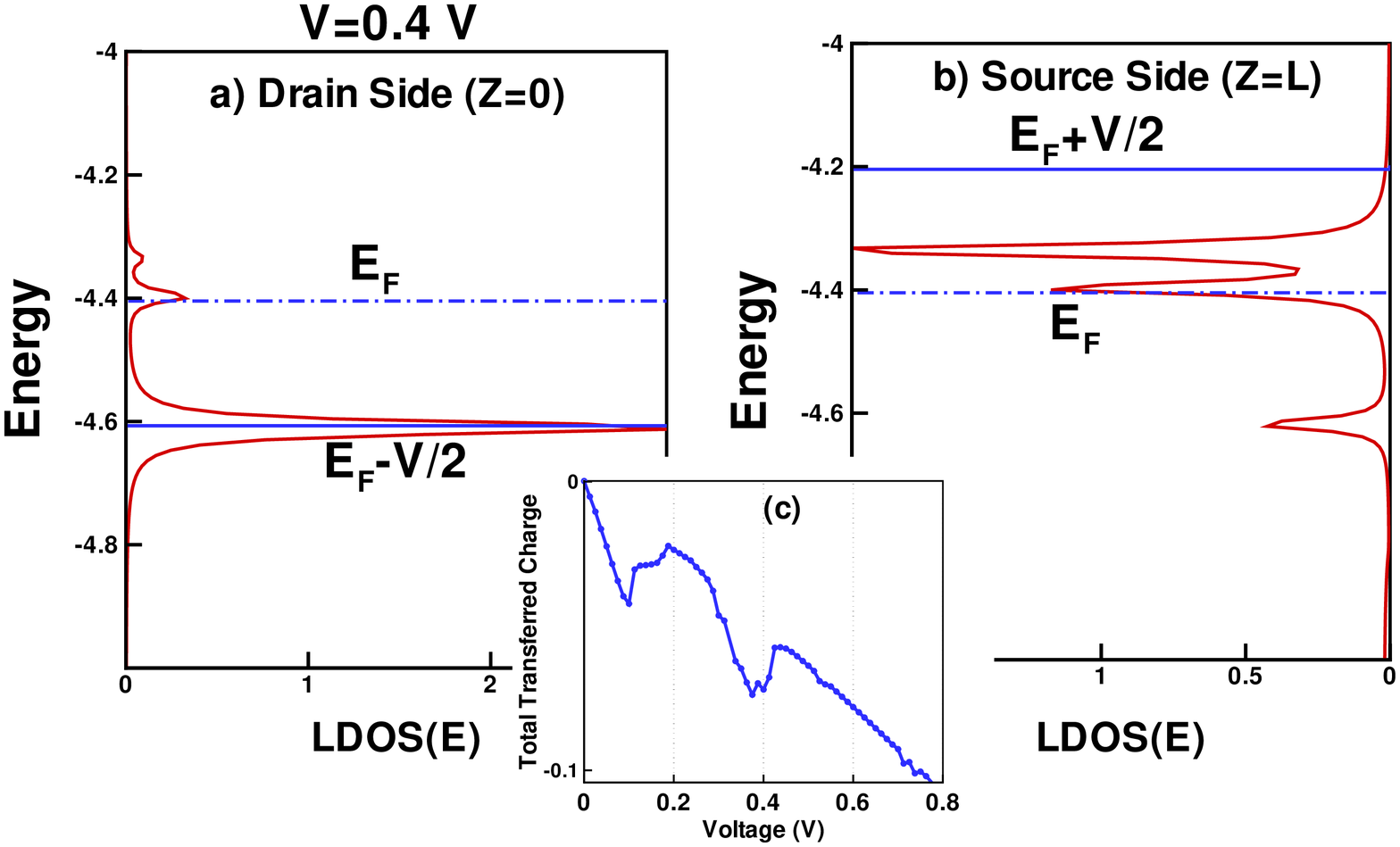}
\caption{(Color online) Surface density of states of a) the source and b) drain
electrodes on z=0 and z=L shows an asymmetric coupling to the
chain. In this case, the chain contains 4 atoms between the two
Graphene tips. c) Total charge depletion ($\delta n$) of the
central region versus applied bias.} \label{SDOS} \efig
%---------------------------------------

The strong reduction in the transmission arises from the
localization phenomenon which occurs due to the sharp linear
potential drop near the source tip. The onsite energies are most
negative on the left side while they are most positive on the
right side of the source tip (atoms located on $20$ and $25
A^{\circ}$ on Fig. \ref{Charge_potential-GRPH}). Therefore the
LDOS of the left side atoms is large at low energies, whereas
that of the right side atoms becomes large at high energies. This
situation is very similar to an ionic bond with a large onsite
energy difference. The bonding and antibonding eigenstates become
farther separated (compared to when onsite energies were equal),
and this causes transfer of charge to the low energy site, and
enhanced localization of orbitals on the sites due to the large
electric field present.

The upper half of each curve in Fig. \ref{LDOS-COND-GR} shows LDOS
on the left and right side atoms of the source tip at voltages of
the peak and valley of current. It was checked from the LDOS data
that states with higher energies become localized on site  labeled
by 25 \AA , while lower energy states become localized on the
left atoms of the source tip (site labeled by 20 \AA). Therefore
the product LDOS at these two sites is reduced with increasing
bias, due to a reduced overlap, leading to a decrease in $T(E)$
according to eq.\ref{transmission}. In Fig. \ref{LDOS-COND-GR}
and for a bias voltage of 0.55 V, strong localization occurs at
$E=-4.25 eV$ where the transmission is also reduced. The lower
half of the curves in Fig. \ref{LDOS-COND-GR}  shows that the
transmission closely follows the product of LDOS of the left and
right atoms (atoms located on 20$A^{\circ}$ and 25$A^{\circ}$) of
the source tip. By increasing the bias from the current-peak to
current-valley, states with higher energies become localized on
the right side of the source tip. So the overlap of LDOS's on the
ends of the source tip is reduced. As a result, their product
which is proportional to the transmission decreases. If these
localized states fall in the integration window of current,
transmission as well as current reduction occurs.

%------------- Fig.6 ----------------
\bfig
\includegraphics [width=7 cm] {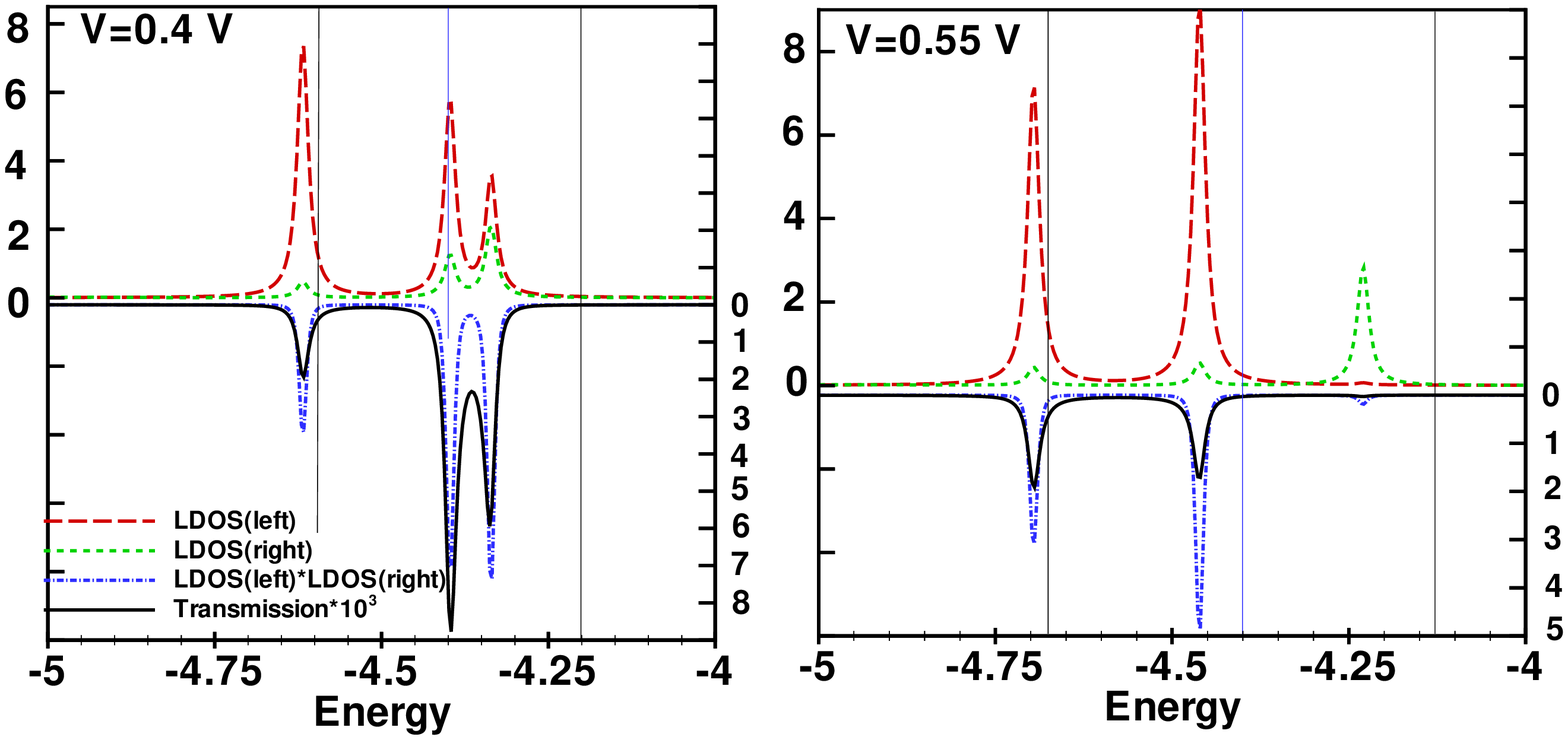}
\caption{(Color online) Local density of states on the first (atom located on
$20A^{\circ}$, long dashed line) and last (atom located on
25$A^{\circ}$, dashed line) atoms of the source tip is plotted in
upper half of the graphs. Their product (dotted line) is compared
with the transmission (solid line) in lower half of the graphs.
Voltages are at the current-peak (0.4 V, left) and current-valley (0.55
V, right). The chain connected to the Graphene tips contains 4 atoms.
Vertical lines show the Fermi level and the integration window.
For comparison with LDOS products, transmission is shown $10^3$
times larger.}\label{LDOS-COND-GR}
 \efig
%--------------------------------------

\section{Conclusion}
In conclusion, for observation of NDR, although the presence of
sharp features in the density of states located on the sharp tip
apexes and their localization is required, the enhancing factor
for localization is the charge depletion of the molecule as the
bias is increased. Asymmetric potential profile which shows a
sharp potential drop in the source side of the molecule, arises
from the asymmetry in the LDOS of electrodes connected to the
molecule. The asymmetry in LDOS's causes different amounts of
charge flow from the molecule to the drain and source electrodes,
respectively. The weak screening of the potential due to the
depleted charge causes a larger potential drop on the source
side. However, the potential on the drain side varies weakly and
remains almost flat. Because of the potential drop in the source
tip, states with higher energy become localized on the sites with
higher potentials (right side of the source tip), and states with
lower energy become localized on the sites with lower potentials
(left side of the source tip), similar to an ionic bond. The
charge depletion and potential drop are intensified in the source
tip as the applied voltage is increased. This results in a more
effective localization of states. Localization causes a reduction
in the overlap of the LDOS's on the ends of the source tip and a
subsequent reduction in the transmission and current.
\section{Appendix}
\subsection{The variational Method}
To find the effect of image charges, we need to impose the
Dirichlet boundary condition $V=0$ at the two left and right
electrode planes. Instead of solving Poisson's equation, we
postulate the electrostatic Green's function of
Eq.(\ref{hamiltonian}) to be:

\beq V(\vec{r}_i,\vec{r}_j)=\left\{\begin{array}{ll}
 \vspace{0.3 cm}
U(\vec{r}_i,\vec{r}_j)-U(\vec{r}_R,\vec{r}_j)&z_i
>z_j
\\ \vspace{0.3 cm}
U(\vec{r}_i,\vec{r}_j)-\frac{U(\vec{r}_R,\vec{r}_j)+U(\vec{r}_L,\vec{r}_j)}{2}&z_i
=z_j
\\\vspace{0.3 cm}
U(\vec{r}_i,\vec{r}_j)-U(\vec{r}_L,\vec{r}_j)&z_i<z_j
\end{array}\right.\label{variational}
 \eeq

where $\vec{r}_R$ and $\vec{r}_L$ show the positions of the atomic
layers located in the right and left contact surfaces,
respectively. Although this function is not the exact solution of
Poisson's equation, it has the correct limits for $r_i$ on the
boundary surfaces, where it is equal to zero by construction. It
is therefore a reasonable solution in a variational sense, though
here we are not varying any parameter to optimize the solution.
In this method, we postulate that the image charges potential on
the test charge plane ($z_i=z_j$) to be as an interpolation of
the left and right solutions in Eq.(\ref{variational}). The
kernel used for the coulombic function $U$ has been chosen to be
as the OK model in Eq.(\ref{OK}).

\subsection{Numerical Method of Images}

\bfig
\includegraphics[width=8 cm]{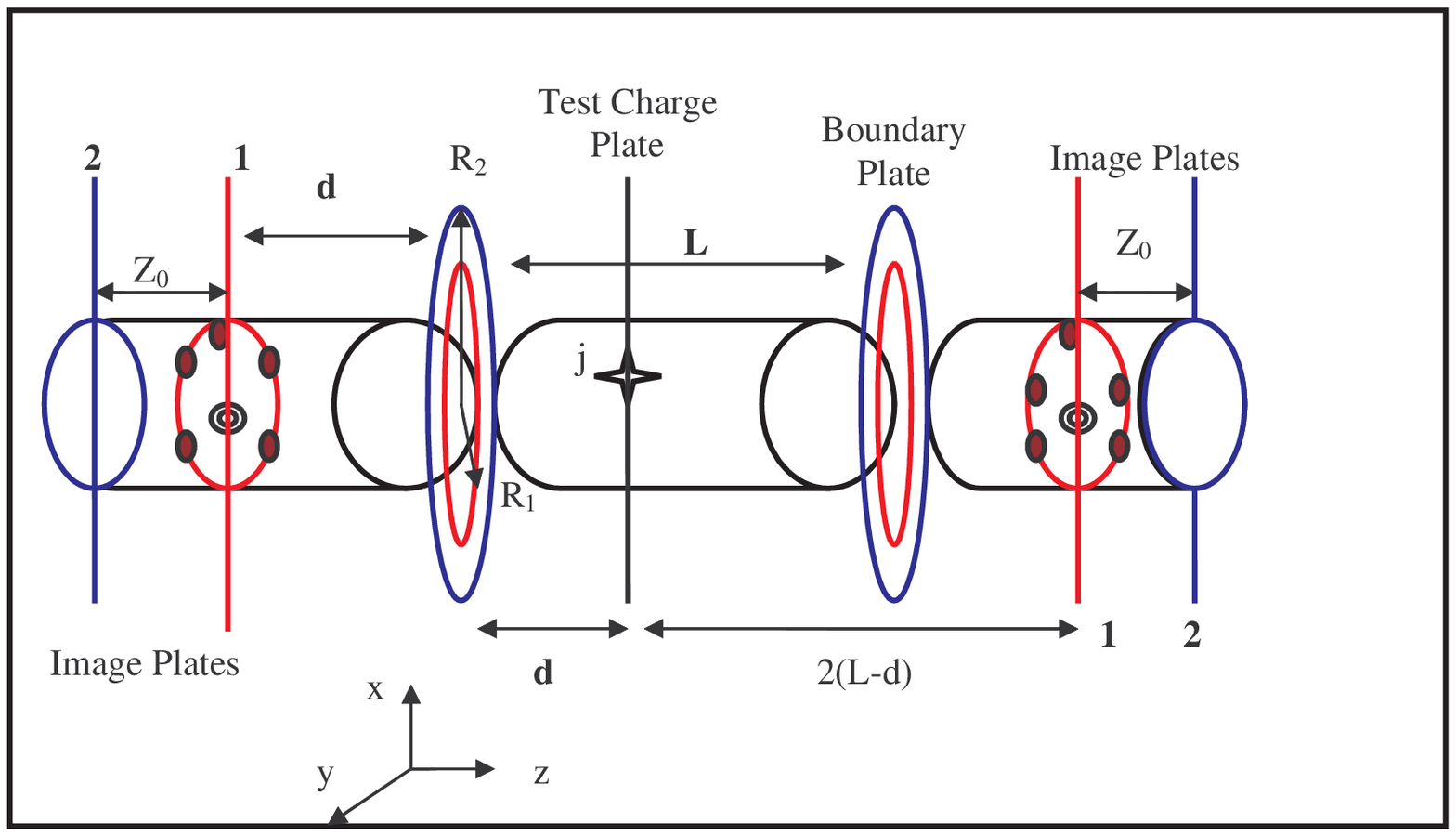}
\caption{In the image charges method, test charge induces some
image rings just behind the boundary surface.
}\label{image-method} \efig

The straightforward way for providing an electrostatic Green's
function which satisfies Dirichlet boundary condition, is to use
image charges. Image charges can be put on fictitious planes just
behind the plane on which we want the potential to be zero. Note
that the choice of their location or charge is not unique.

Since the potential on the boundary surfaces must be zero, one
can find the image charges, if their location is fixed, by
solving a system of the linear equations. For a test charge
located on a molecular site $\vec{r}_j$, one has to solve the set
of linear equations which are equal in number to the number of
boundary constraints. The constraints leading to a linear system
are as follows:

\beqnar
V(\vec{r}_i,\vec{r}_j)&=&U(\vec{r}_i,\vec{r}_j)+\sum_{k=1}^{n_{img}}q^{j}_k
U(\vec{r}_i,\vec{p}^{j}_{k}) \\
V(\vec{r}_L,\vec{r}_j)&=&V(\vec{r}_R,\vec{r}_j)=0 \label{images}
\eeqnar

where $\vec{r}_{i}$ is the field point and $\vec{r}_{j}$ is the
source point, with its images being of charge $q^{j}_k$ and
located at $\vec{p}^{j}_{k}$. For a given test charge location,
the number of images $n_{img}$ we need depends on the number of
points (constraints) on the boundaries, at which one wants the
potential to be zero.

 \bfig
\includegraphics [width=8 cm]{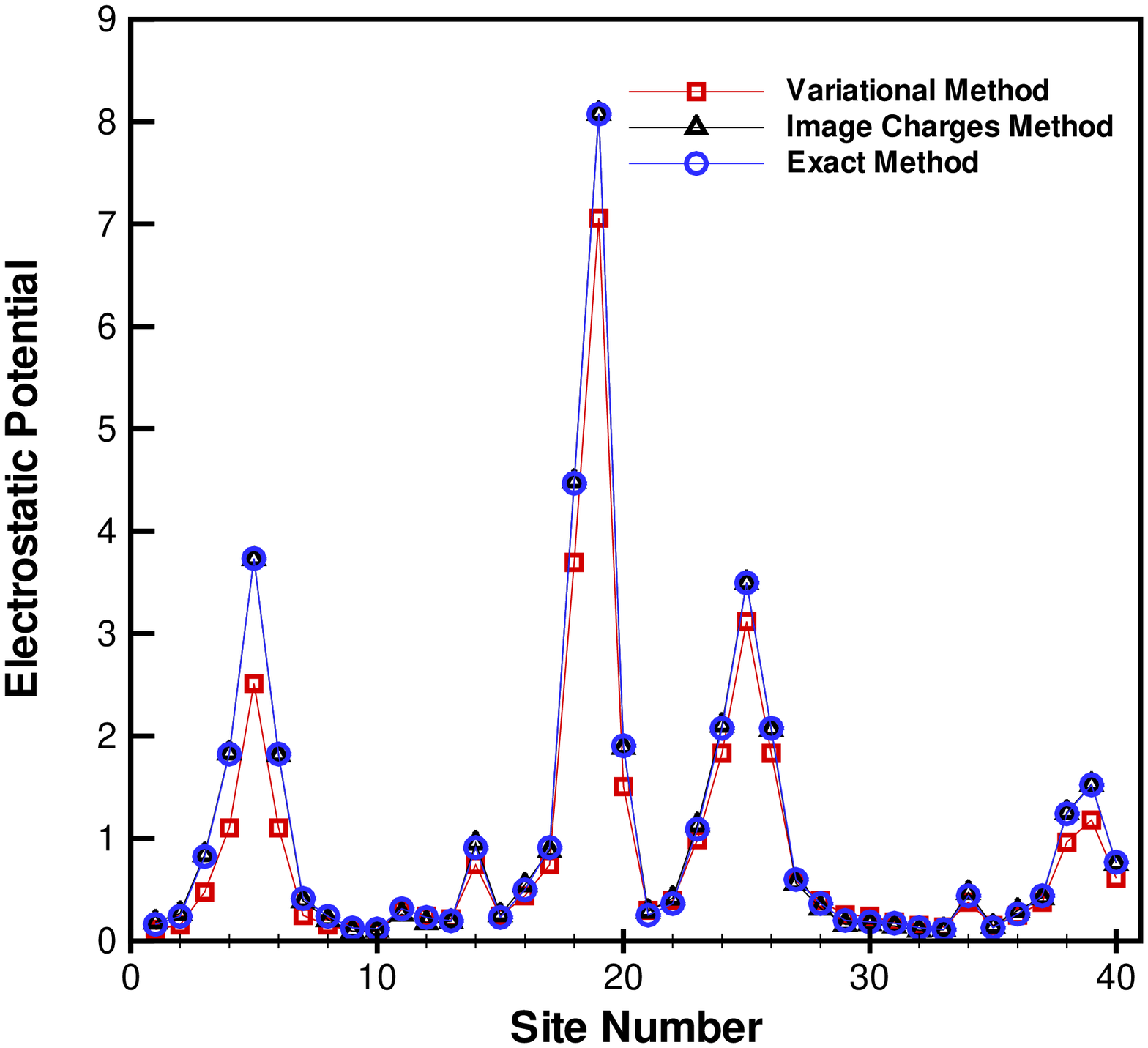}
\caption{Comparison of three methods for the calculation of the
electrostatic Green function. The sample is a (5,5) nanotube
which has 4 rings (40 atoms) in the middle part. The test charge
is set on the site number 19. The Hubbard term is considered to
be $U_H=11.3$. Numerical calculation of the image charges method
has been done by $n=20$ and $z_0=2$.}\label{comparison} \efig

As an example, Fig.(\ref{image-method}) shows a nanotube and the
position of its contacts and image charges. In this model, all
image rings are placed behind the first image plane marked by
number $1$. The first image charge planes which are the reflected
planes from the contact surfaces, are located at $z=-d$ and
$z=2L-d$, where $d$ is the distance of the plane which includes
the test charge from the left contact surface. The distance of
image planes from each other is considered to be a constant value
$z_0$. The number of image planes is equal to the number of
boundary rings ($n$). It is supposed that the number of sites on
an image ring is the same as the boundaries and nanotube rings.
In this case, cylindrical symmetry of the images and boundaries
sites is important to produce a smooth potential at the
boundaries.

Fig.(\ref{comparison}) shows a {\it comparison} between these
three methods. A good correspondence can be observed between the
potential of image charges method and the exact method. They
differ by only 2 percent, while they have about $20$ percent
difference with the variational method. However, the advantage of
the variational method is its simplicity for application on any
structure, while the position and values of image charges depend
on the structural symmetries.


\begin{thebibliography}{99}
\bibitem{ndr-gf}
This is due to either a symmetry mismatch between left and right
wavefunctions(see for example refs.
[\onlinecite{farajian,orb-match}], or an exponential decay versus
length of the GF due to energy non-alignment (for example in RTD
diodes\cite{esaki,esaki-rtd,nanolett1}).
\end{thebibliography}
\end{document}